# Gravity from Transactions: Fulfilling the Entropic Gravity Program

A. Schlatter* and R. E. Kastner**

*The Quantum Institute, NY, USA;
**University of Maryland, College Park; The Quantum Institute; rkastner@umd.edu

**Abstract:** This is a comprehensive review of new developments in entropic gravity in light of the Relativistic Transactional Interpretation (RTI). A transactional approach to spacetime events can give rise in a natural way to entropic gravity (in the way originally proposed by Erik Verlinde) while also overcoming extant objections to that research program. The theory also naturally gives rise to a Cosmological Constant and to Modified Newtonian Dynamics (MOND) and thus provides a physical explanation for the phenomena historically attributed to "dark energy" and "dark matter".

**Keywords:** transaction; quantum-event; space-time; emerging gravity; causal sets; Einstein equation; cosmological constant; de Sitter space; Schwarzschild space; MOND.

## 1. Introduction and Background

In previous publications, one of us (R.K.) has been developing the Transactional Interpretation into the relativistic domain (e.g. [1 − 3] ; now called RTI), while the other (A.S.) has been applying it to the problem of quantum gravity [4 − 6]. In the present work, we review some key results and show how RTI can provide a "missing link" that greatly reinforces the program of entropic gravity pioneered by E. Verlinde [7] by giving a specific quantum-level account of the entropic effects that is immune to extant objections. In particular, entropic gravity is often thought to arise from "quantum entanglement", but this does not provide a clear connection between the quantum level and that of spacetime, and is also subject to criticism based on the lack of appropriate decoherence effects. In contrast, since transactions automatically trigger decoherence [2], in the present approach these weaknesses are remedied. In what follows, we review key developments and discuss in specific terms the role that the transactional picture plays in deriving entropic gravity and in resolving current challenges for this program.

We first introduce some key terms that will be used to make contact with the existing literature and to develop the current proposal:
1. quantum substratum: pre-spatiotemporal domain housing quantum systems with non-vanishing rest mass
2. spacetime: a storage device for measurement results and their structural connections (i.e., on-shell photons)
3. matter: quantum system with non-vanishing rest mass
4. holographic principle: description of a spacetime region can be thought of as encoded on a lower-dimensional boundary to the region [8]
5. emitter: an elementary bound matter system in an excited internal state (e.g., atom or molecule)
6. absorber: an elementary bound matter system subject to excitation of its internal states (e.g., atom or molecule)
7. mass (m) M: a fermionic bound state in its local rest frame with rest mass, (m) M, possibly composed of N elementary systems
8. transaction: general term for a non-unitary process ultimately leading to transfer of energy (and other conserved quantities) in the direct-action (absorber) theory of fields [9]
9. actualized transaction: the transfer of a real photon from one emitter to one absorber, generating an emission and absorption event and their invariant connection (null interval). The latter constitute elements of spacetime.

The above definitions indicate that the current work views spacetime as non-fundamental, which breaks with the traditional view of spacetime as a container or background for all physically real entities. Instead, we adopt a view that the covariant construct described by general relativity is emergent by way of a specific non-unitary transformation corresponding to quantum measurement, reviewed below. One can also see in points

5. and 6. above that there is some overlap between an "emitter" and an "absorber". These roles can depend on the relation between the elementary mass (m) and other elementary masses. While a ground state atom can only act as an absorber, an excited atom could either emit or absorb depending on the availability of other atoms in higher or lower states separated by energies $h\nu$ corresponding to those of the given atom.

*1.1 Relativistic Transactional Interpretation as an account of the emergence of spacetime events and their structural connections*

The Relativistic Transactional Interpretation (RTI) suggests that the spacetime manifold does not exist as a primary ontological structure, but supervenes on specific sorts of interactions among quantum systems--i.e. *transactions*, which will be briefly reviewed here. Quantitative details of this formulation are given in [1]. RTI predicts the emergence of spacetime events from a more fundamental *quantum substratum*, such that an actualized transaction gives rise to an emission event, an absorption event, and the invariant null interval connecting them, which is constituted by the real, transferred photon. Thus, this is a "growing universe" type of picture.[1] In what follows, we show how the concept of a transaction, as elaborated in RTI, constitutes the crucial "missing link" between the quantum level and the emergent spacetime level that allows for a fully consistent and seamless formulation of entropic gravity.

RTI is based on the fully quantum form of the direct-action or "absorber" theory of fields ([1], *Chapter* 5). In this picture, unitary quantum processes and interactions include the mediation of force through direct connections via the time-symmetric propagator, which is oblivious to temporal orientation. Accordingly, such interactions are not spacetime processes, but act "behind the scenes", as physical possibilities (cf. Kastner, Kauffman and Epperson [10] ), to steer the probabilities for the actualizations of spacetime events (the latter involving a non-unitary process to be discussed further below). The systems involved, such as elementary particles and atoms, are subject to quantum entanglement and accordingly are described by Hilbert Space states. As elements of a vector space of 3N spatial dimensions with complex amplitudes, such states (and accordingly the entities they describe) are clearly not commensurable with the structure of spacetime. Thus, the domain of the systems and their interactions cannot be 3+1 spacetime and accordingly, as mentioned above, we identify it as an extra-spatiotemporal quantum substratum (QS). This is actually consistent with standard general relativity, since it is well known that the curvature associated with any mass source is nonvanishing. Were the mass source within the spacetime construct, the curvature would have to vanish at the mass.[2] Another type of unitary process in the QS is the Schrödinger evolution of fermionic matter, such as electrons, or bound states such as atoms. One can attribute an internal "clock" to such systems, corresponding to their spin [11] which serves to define their proper time. The Schrödinger evolution of material systems defines inertial frames for the emergent spacetime, though they themselves are not features of spacetime.

As alluded to above, one requires a non-unitary interaction called a "transaction" in order to generate spacetime events and their connective structure. The non-unitarity is inherent in the direct-action picture of fields (but is obscured in the standard formulation), and can be quantified in terms of decay rates ([1], *Chapter* 5). The non-unitary interaction is triggered by what is called "absorber response" in the non-relativistic form of the theory, but at the relativistic level (RTI) is actually a mutual interaction among emitters and absorbers, which is called the "NU-interaction" (for "non-unitary interaction"). A basic requirement for the NU-interaction is the satisfaction of conservation laws. It is only through the NU-interaction that a real, on-shell photon can be created ([1], *Chapter* 5). Once a real photon is created, energy is transferred from the emitter to one of the responding absorbers in a radiative process. This process establishes a "link" of spacetime constituted by the emission event, the absorption event, and the transferred photon. These links are invariant null intervals. Thus, spacetime is constituted solely by invariant events and their invariant null connections. In this context, the "field" is the structured set of events. The spacetime construct itself is fully invariant, and it is only descriptions, relative to internal inertial frames, that are non-invariant. In this regard, it is vital not to conflate the map (coordinate-dependent descriptions such as "spacetime diagrams'') with the actual territory (the invariant events and their null connections).

---

[1] In this picture the "Big Bang" was simply the initial emergence of the structured set of events that we call "spacetime", and which continues to grow. Transactions are also the reason for the Hubble expansion, as we show in section 3.

[2] Thus, the common notion that "everything physical exists in spacetime" is an uncritical folk belief that should be discarded.

It is also important to note that the events are not to be identified with the material systems (e.g., atoms) giving rise to them. That is, the systems themselves are never part of the spacetime manifold nor are they within it. It is only their *activities*--the events--that are elements of the spacetime manifold, as are the photonic connections between them, which establish the structure of the manifold. In common parlance one refers to a spacetime point being "occupied'' by matter, but in our picture that is really a shorthand for the idea that a material system has become associated with that point (understood as simply an element of a particular reference frame) via an actualized event. (Of course, given the spreading of wave packets, such a system will not remain indefinitely localized.)

*1.2 The stress-energy tensor describes classical radiation fields, not individual photons*

We deal first with a possible concern. It is noted above that photons are radiation. It is often stated that radiation is a source of the field, but this statement requires crucial disambiguation in the quantum context, and especially in the light of RTI. This is because individual photons (Fock states) are not sources of the gravitational field, even though they reflect "curvature'' of the metric due to matter sources. Consider the stress-energy tensor $T_{\mu\nu}$. The standard account notes that the contribution, to the energy density $T_{00}$ for radiation is with specific constants $\varepsilon_0, \mu_0$ and $\vec{E}, \vec{B}$ denoting the electric and magnetic fields

$$T_{00} = \frac{1}{2c^2}\left(\frac{1}{\varepsilon_0}|\vec{E}|^2 + \mu_0|\vec{B}|^2\right). \tag{1}$$

Other components of the stress enrgy tensor involve the momentum or flux, which is expressed through Poynting's vector

$$\vec{S} = \varepsilon_0 c^2 \vec{E} \times \vec{B}. \tag{2}$$

However, the classical fields $\vec{E}$ and $\vec{B}$ vanish for Fock states, so individual photons do not contribute to $T_{\mu\nu}$. The way we obtain non-vanishing classical-level fields by reference to quantum theory is through so-called "coherent states" $|\alpha\rangle$, which actually comprise an indefinite number of photons:

$$|\alpha\rangle = e^{-\frac{|\alpha|^2}{2}} \sum_{n=0}^{\infty} \frac{\alpha^n}{\sqrt{n!}}, \alpha \epsilon \,\mathbb{C}. \tag{3}$$

According to RTI, however, such states do not really describe individual photons. Rather, they describe collective emitter and absorber states that give rise to probabilities of photon detection. It is these detections of actualized, well-defined numbers of photons that collectively trace out the amplitude of the classical field. (For further details, see [12]). Thus, the individual photons that connect emitters and absorbers in actualized transactions, and thereby serve as the structural connections of spacetime through the invariant null intervals that they establish, do not themselves constitute sources of the gravitational field--i.e., spacetime. Rather, they are structural components of spacetime.[3] The null intervals are one-dimensional constituents of space-time, which emerge by interacting matter. The notion of micro-constituens of space-time in general is discussed with a view to the vast literature in e.g. [13].

*1.3 The role of spacetime*

---

[3] This is consistent with the observation that while the expectation values of the $E$ and $B$ field operators are zero in a Fock state, their variances are nonzero (and maximal). The large variance of the E and B field operators is a reflection of the completely uncertain phase of the Fock state, which assures that new spacetime events are added in a Poissonian manner. The latter assures covariance of the coordinate-level description (see, e.g., Sorkin 2003 [16] ).

As noted above, we do not take spacetime as a "container", so care is needed regarding terminology like "spacetime region" or "masses at distance $R > 0$". The spacetime manifold is simply a structured set of emission-and absorption events. This was essentially noted by Einstein, who remarked:"There is no such thing as empty space, i.e., a space without a field. Spacetime does not exist on its own, but only as a structural quality of the field"[14].

In our proposal, the term "field," as used by Einstein, refers to the structured set of events actualized by way of transactions. This picture is harmonious with the analysis provided by Westman and Sonego (2009), who provide further insights into Einstein's fundamentally relational view of spacetime[4]. The events constitutive of the field, which Einstein referred to as "point-coincidences," are essentially the emission and absorption events. Westman and Sonego [15], although they use the term "field" somewhat differently, note the crucial distinction between invariance and covariance, which applies to our proposal as follows: the actualized set of events is invariant (frame-independent), while the description in terms of the spacetime parameters is covariant (frame-dependent).

There is an a priori spatial and temporal intuition, no doubt, and we associate to physical systems spacetime parameters. But it is the exchange of photons which actualizes the events constitutive of the invariant spacetime manifold and connects emissions and absorptions by generating a relation between them in the form of a spacetime null-interval. The set of possible events together with their metric structure can be modeled as points of a four dimensional manifold and their distances. So a "spacetime region" simply maps the activities of masses that generate the events, but who exist beyond the spacetime construct associated with them. Objects of geometric calculations are elements of this idealized, continuous and in itself static manifold-model of possible events, out of which grows by transactions a discrete "galaxy" of actualized events–the latter being the actual, emergent spacetime manifold. We treat the pre-emergent manifold of possible events as a map or "blueprint" and even assign properties to its elements (points, regions), in order to formulate relations between the actualized events. We will see in section 3 that the most general covariant description compatible with such a set of events is in terms of a Lorentz-manifold.

It is thus, in the context of RTI, that the holographic principle acquires a new interpretation: the holographic screen can be thought of as the boundary between the quantum substratum and the newly emergent region of spacetime, in the sense above. This picture is harmonious with the original idea in Verlinde [7] in which he noted that such a screen can be thought of as containing data about what is on the not-yet-emergent side of it.

In the sequel we will show in section 2 that transactions are the physical basis of the entropy, which is used in [7] to derive Newtonian gravity. We will further show in section 3 how the exchanged photons can serve as clocks, naturally inbuilt to events, whose gauging leads most generally to Einstein's equations on a Lorentz-manifold, including a cosmological term. The synchronization of these light-clocks from the perspective of different observers in specific spacetimes further leads to the correction of Newtonian gravity as calculated in MOND, [18, 19], a fact which we show in section 4. Finally in section 5 we state some final conclusions.

**2. Gravitational Force from Entropic Considerations based on Transactions**

More formally, we can arrive at the basic entropic argument and its relation to gravitation, as outlined above, from information-theoretic considerations and the mechanism of transactions.

*2.1 Localization and entropy*

By a transaction an absorber in energy-groundstate $|E_0\rangle$ measures a photon four-momentum $|k_j\rangle, j\epsilon\mathbb{N}$, [5] by assuming an energy-pointer state $|E_k\rangle, k\epsilon\mathbb{N}$. The corresponding probabilities are $p_k = |\langle E_k|E_0\rangle|^2$. By the transfer of the photon, the transaction also localizes the absorption with spatial component $x_j = x(\vec{k}_j)\epsilon\mathbb{R}^3$ [6].

---

[4] For background on relational vs. substantival theories of spacetime, see Friedman [17].

[5] We assume that the photon four-momentum space $\mathcal{P}$ is coarse-grained, $\mathcal{P} = \{|k_j\rangle\}_{j\epsilon\mathbb{N}}$. This can be expected in a direct action theory like RTI.

[6] Emission is also localized and, due to the transaction which establishes a spacetime distance, we can work in a joint coordinate system of both emitter and absorber. Connected components of events lie thus in one chart of the coordinate system.

There is no covariant space-operator and non-relativistic position-and energy-operators do not commute. So the localization is not a "measurement" but a consequence of the space-time interval, which emerges between emitter and absorber by a transaction. There is a set of corresponding position-probabilities

$$\tilde{p}_{kj} = |\langle E_k | x_j \rangle|^2. \tag{4}$$

The corresponding (Shannon-) information is

$$\tilde{I}_k = -\sum_j \tilde{p}_{kj} ln(\tilde{p}_{kj}). \tag{5}$$

It is formally possible to transform $\tilde{I}_k$ into (thermodynamic) entropy $\tilde{S}_k$ by multiplication with the Boltzmann constant $k_B$.[7] $\tilde{S}_k$ is independent of the probabilities $p_k$ of the energy-states $|E_k\rangle$, but depends, of course, on the physical systems involved and, if there are large numbers of emitters and absorbers involved[8], some generic-quantity is necessary. This can be achieved by changing perspective and by asking, whether there is some measure for the information content of space independently of specific material systems. It turns out that a corresponding measure can naturally be defined and it is hence possible to allocate information content to spatial regions per se. This information actually leads to the entropic gravity in [7].

*2.2 Spatial information*

We assume to work in a local inertial frame throughout the following exposition. Let there be a bounded region $\Omega \subset \mathbb{R}^3$ on a spatial hyperplane and a partition by balls

$$\mathcal{B} = \{B_{\varepsilon_n}(x_n)\}_{x_n \epsilon \Omega, \varepsilon_n > 0}, \cup_{x_n} B_{\varepsilon_n}(x_n) = \Omega. \tag{6}$$

Relative to the partition $\mathcal{B}$, a position-information can be attributed to a quantum system in terms of square-integrable functions over $\Omega$, $\psi(x) \epsilon L^2(\Omega)$, by

$$I^{\mathcal{B}}(\psi) = -\sum_{x_n \epsilon \Omega} p_{x_n} ln(p_{x_n}), \qquad p_{x_n} = \int_{B_{\varepsilon_n}(x_n)} |\psi(x)|^2 dx. \tag{7}$$

By multiplication with the Boltzmann constant, $k_B$, we get

$$S^{\mathcal{B}}(\psi) = I^{\mathcal{B}}(\psi) k_B. \tag{8}$$

We can ask, whether it is possible to take a different perspective and attribute information not to material systems, but to regions or, idealized, single points $x_0 \epsilon \mathbb{R}^3$. A point, $x_0 \epsilon \Omega$, can empirically be associated with matter or not and hence represents in this sense one bit of information. Given a single physical system described by $\psi(x) \epsilon L^2(\Omega)$, we can therefore state that the information of the one bit, $x_0 \epsilon \Omega$, with respect to $\psi(x)$ and the partition $\mathcal{B}$ (6)[9] is

$$I_\psi^{\mathcal{B}}(x_0) = -[p_{x_0} ln(p_{x_0}) + (1 - p_{x_0}) ln(1 - p_{x_0})]. \tag{9}$$

---

[7] While formally correct, the physical validity of this identification is still debated [20]. Among the authors R.K. has reservations, while A.S. supports it as the most general form of Boltzmann's law.

[8] For instance in the case of macroscopic systems.

[9] We pick the ball $B_{\varepsilon_{\tilde{n}}}(x_{\tilde{n}})$ with $x_0 \epsilon B_{\varepsilon_{\tilde{n}}}(x_{\tilde{n}})$ and minimal $|x_0 - x_{\tilde{n}}|$.

To find a generic definition we have to account for all possible partitions $\mathcal{B}$, which amounts to taking into account all probabilities, $0 \leq p_{x_0} \leq 1$. Since we always find some bounded $\Omega \subset \mathbb{R}^3$ with $x_0 \epsilon \Omega$, we can define the information $I(x_0)$, $x_0 \epsilon \mathbb{R}^3$, by

$$ \tag{10}$$

Evidently, (10) is not only independent of a chosen partition $\mathcal{B}$, but also of the particular state $\psi(x)$. While the choice of a particular $\mathcal{B}$ is, of course, frame-dependent, the described process will lead to the definition of $I(x_0)$ by equation (10) in every local inertial frame. By (8) we arrive at the corresponding one-bit entropy

$$I(x_0)k_B = \frac{1}{2}k_B. \tag{11}$$

Because localization is physically limited to a Planck volume, we rescale the entropy (11) by a factor of $4\pi$, to finally get

$$S(x_0) = 2\pi k_B. \tag{12}$$

*2.3 Transactions and a holographic principle*

Assume there is a transaction between two physical systems through the exchange of a real (on-shell) photon. A transaction breaks the unitary symmetry of the interaction and localizes absorber and emitter. If it takes the photon a time interval $\Delta t_R = \frac{R}{c}$ from the emitter to the absorber, then, next to the time interval $\Delta t_R$, there is the spatial distance $R > 0$ being generated by the process. For symmetry reasons[10], the possible locations of the absorber lie equiprobably on the sphere $S_R$. We can define in analogy to (10) and (12) the total entropy of the sphere $S(S_R)$ around the emitter by

$$S(S_R) = 4\pi k_B I(S_R) = 2\pi k_B N_R, \tag{13}$$

where $N_R$ is just the number of bits on the sphere. For this number we set, with $l_P = \sqrt{\frac{G\hbar}{c^3}}$ and $A_R = 4\pi R^2$,

$$N_R = \frac{A_R}{4\pi l_P^2} = \frac{R^2 c^3}{G\hbar}.\text{[11]} \tag{14}$$

So we get

$$S(S_R) = 2\pi k_B \frac{A_R}{4\pi l_P^2} = k_B \frac{A_R}{2 l_P^2}.\text{[12]} \tag{15}$$

Remember that expression (13) is the sum of all entropies $S(x)$, $x \epsilon S_R$, which in turn are calculated over all possible local probabilities, $0 \leq p_x \leq 1$, $x \epsilon S_R$. So $S(S_R)$ can be considered to be the entropy contained in all possible absorptions on the sphere $S_R$.

Let us assume that in a concrete physical situation the fields defined over the ball $B_R$, $\psi(x) \epsilon L^2(B_R)$, have a total mass $M$. We then know the total energy to be $E_{tot} = Mc^2$ and derive from (15) and the thermodynamic relation $E = S \cdot T$ by the thermodynamic equivalence principle the existence of a formal temperature $T = T(M, R)$ satisfying

---

[10] Photons are emitted in all spatial directions with equal probability.

[11] Note that equation (14) might serve as a definition for the constant $G$ [7].

[12] We can multiply the information-entropy $I(x)$ by the "number" of bits on the sphere, because $I(x)$ is the average information over all partitions containing the point $x$ and is hence independent of a specific coverage $\mathcal{B}$ of the sphere. In order to count the number of bits on the sphere, $S_R$, we chose the local surface-measure $d\sigma(x) = (4\pi l_P^2)^{-1} d^2(x)$ and get $S(S_R) = \int_{S_R} k_B I(x) \, d\sigma(x) = k_B \frac{A_R}{2 l_P^2}$.

$$Mc^2 = S(S_R) \cdot T(M,R) = k_B \frac{A_R}{2l_P^2} T(M,R). \tag{16}$$

By (11) we get for $T(M,R)$, with $g_R = \frac{GM}{R^2}$, the expression

$$T(M,R) = \frac{\hbar MG}{2\pi c k_B R^2} = \frac{\hbar g_R}{2\pi c k_B}. \tag{17}$$

We are now ready for the final step.

### 2.4 Entropic force

We are now in a position to reconstruct the argument in [7]. Assume that a particle of mass $m$ enters into a transaction with a system of total mass $M$, $m \ll M$, a process which localizes the particle at a point $x_0 \epsilon S_R$ for some $R > 0$. This process makes position-information (12) available and consequently by the 2nd law, there is a (minimal) amount of entropy added to the thermal environment on the surface of $\Delta S(x_0) = 4\pi k_B I(x_0)$.[13] We account for the fact that a particle is not really point-like, but still structureless, by using its (reduced) Compton radius $\lambda_C = \frac{\hbar}{mc}$ and postulate that the full information is part of the space-like surface $S_R$ only, if the particle is at a Compton-distance from it and that the information $I_{\Delta x}(x_0)$ decreases linearly, if it is getting closer to the surface, (i.e. [7], [21])

$$I_{\Delta x}(x_0) = \frac{1}{2} \frac{mc}{\hbar} \cdot \Delta x, \quad 0 \leq \Delta x \leq \lambda_C. \tag{18}$$

Associated with the entropy difference there is by (17) a (minimal) amount of energy $E_{\Delta S(x_0)}$ given by

$$E_{\Delta S(x_0)} = 4\pi k_B I_{\Delta x}(x_0) T(M,R). \tag{19}$$

The definition of energy as work, i.e. force along a path, leads to a corresponding force $F_G$ by

$$E_{\Delta S(x_0)} = F_G \cdot \Delta x. \tag{20}$$

By (17) and (18) we get from (20)

$$F_G \cdot \Delta x = 2\pi k_B \frac{mc}{\hbar} \frac{\hbar MG}{2\pi c k_B R^2} \cdot \Delta x, \tag{21}$$

and finally the expression for $F_G$

$$F_G = m \cdot g(M,R) = m \frac{GM}{R^2}. \tag{22}$$

The force $F_G$ is attractive, since the entropy-gradient $\Delta S(x_0)$ points to the surface $S_R$. The above derivation of $F_G$ has shown that the concept of spatial information and transactions leads, together with the thermodynamic equivalence principle (16), to the existence of gravity as an entropic force, emerging from the coming-into-being of empirical reality (22).[14]

---

[13] As explained in 2.1, entropy is understood as unavailable position-information, which is calculated as an average (10,11,12) over all possible states. Transactions make this information available, a fact which must be compensated by the second law. The localization acts as a constraint force whose magnitude is equal to F$_G$. This recalls that F$_G$ is not a "real force," but appears relative to an inertial frame.

[14] We have worked out the minimal acceleration, based on the minimal dissipation in (19). Generally, we can expect
$a(M,R) \geq g(M,R)$. We will see in paragraph 3 that this possibility is implicitly contained in Einstein's equations.

Once we consider that transactions possess by the exchanged photons naturally inbuilt clocks, which measure the rhythm of becoming by an invariant gauge $c$, the speed of light, the way is open to derive a metric structure of spacetime, locally governed by the Einstein equations [4]. In the process we find the reason for the general validity of the clock-hypothesis.[15] We are going to derive Einstein's equations in this spirit in the next paragraph. The universality of gravity, its propagation at the speed of light and the equivalence of inertial and heavy mass are further immediate corollaries of our insight.

### 3. Einstein Equations

By results in [22,23] it takes a physical system in a pure state $|\psi\rangle$ and in a thermal environment of temperature $T_0$ a time-step of

$$\Delta t = \frac{h}{4 k_B T_0} \tag{23}$$

to flip into an orthogonal and hence distinguishable state $|\psi^\perp\rangle$, $\langle\psi|\psi^\perp\rangle = 0$. This time step is universal and independent of the particular system. The quantity

$$d\tau = \frac{ds}{\Delta t} = \frac{4}{h} k_B T_0 ds \tag{24}$$

measures proper-time $ds$ in units of $\Delta t$. This way, the thermal environment defines the steps of a thermal clock.

We may now assume that spacetime is just locally flat.[16] Like in section 2 let there be a test system with mass $m$ at a sufficiently close distance $R$ from a mass $M \gg m$. Further assume that both systems are instantaneously at relative rest. The test system will by (22) feel an acceleration $g_R$. Using Rindler coordinates, $(\vartheta, \varrho, y, z)$[17], there holds by (24) in the local rest frame at $\varrho = \frac{c^2}{g_R}$ and for $T_{g_R} = T(M, R)$ (17)

$$d\tau = \frac{4}{h} k_B T_{g_R} c d\vartheta. \tag{25}$$

Through multiplying $d\tau$ by the factor $\frac{c}{g_R}$ we get a velocity

$$v_{g_R} = \frac{4}{h} k_B T_{g_R} \frac{c^2}{g_R}. \tag{26}$$

Remember that the acceleration $g_R$ originated from a transaction and the corresponding exchange of a photon. If we gauge the thermal clock (24) by the light-clock, given naturally by the photon[18], we have to gauge (26) by the speed of light

$$\frac{4}{h} k_B T_{g_R} \frac{c^2}{g_R} = \frac{c}{\pi^2}.\,^{19} \tag{27}$$

Solving for the temperature we arrive at

$$T_{g_R} = \frac{\hbar g_R}{2\pi k_B c}. \tag{28}$$

---

[15] Timelike curves on a Lorentz-manifold can be approximated arbitrarily closely by a zig-zag of null curves (i.e. light clocks) [24].

[16] "Locally flat" means approximately Minkowski in small regions around a point $x_0$. Technically this amounts to the existence of geodesic normal coordinates, in which there holds $g_{ab}(x_0) = \eta_{ab}$, $\Gamma^k_{ij}(x_0) = 0$, but generally $\Gamma^k_{ij,\nu}(x_0) \neq 0, 0 \leq i, j, k, \nu \leq 3$.

[17] The line-element in Rindler coordinates is $ds^2 = \left(\frac{\kappa\rho}{c^2}\right)^2 c^2 d\vartheta^2 - d\varrho^2 - dy^2 - dz^2$.

[18] Light signals, sent from one system to an other, are clocks in the sense of Einstein.

[19] The factor $\frac{1}{\pi^2}$ results from going through the same steps leading to equation (23) for a photon-system with $E = \hbar\omega$ [4].

Formula (28) is exactly the Davies-Unruh temperature, which we have already found in equation (17), proving that our approach is consistent and that the thermal clock with temperature $T(M,R)$ is indeed the equivalent of a light-clock. Equation (27) is now the starting point to understand Einstein's equations.

With $E = Mc^2$, $l_P = \sqrt{\frac{G\hbar}{c^3}}$ and $A_R = 4\pi R^2$ we can rewrite equation (16) as

$$g_R A_R = \frac{4\pi G}{c^2} E. \qquad (29)$$

We are interested in the time development of equation (29) and work in a local inertial frame around $M$ in geodesic normal coordinates. With $V_R(t)$ denoting the volume of a small ball of test systems at radius $R(t)$ around $M$, with $R(0) = R$, $\dot{R}(0) = 0$ and $\ddot{R}(0) = g_R$, we can rewrite (26) as [25]

$$\frac{d^2}{dt^2}\Big|_{t=0} V_R(t) = \frac{4\pi G}{c^2} E. \qquad (30)$$

If we introduce the zero-component $T_{00}$ of the energy-momentum tensor $T_{\mu\nu}$, $0 \leq \mu, \nu \leq 3$, via $T_{00} = \lim_{R \to 0} \frac{E_R}{V_R}$ [20], denoting the energy density at the origin, and use the local properties of the Ricci tensor $R_{\mu\nu}$, $0 \leq \mu, \nu \leq 3$, we have [25]

$$\frac{\ddot{V}_R}{V_R}\Big|_{t=0} \xrightarrow{R \to 0} c^2 R_{00}, \qquad (31)$$

i.e. by (29)

$$R_{00} = \frac{4\pi G}{c^4} T_{00}. \qquad (32)$$

So far, we have only made use of the zero-component, i.e. the energy, of the transferred photon. From the transferred three-momentum there arises a pressure $\overline{P}_\gamma$, equivalent to an energy density, which hence enters on the right hand side of equations (29,32). Let $A_i$, $1 \leq i \leq 3$, be small surface elements with $\langle \vec{n}_{A_i} | \vec{e}_j \rangle = 0$, $i \neq j$. We have, with the de Broglie-relation $|\vec{p}| = \frac{h}{R}$ [21], $x_0 = ct$ and $N_R(x_0)$ denoting the number of transactions in volume $V_R$,

$$\overline{P}_\gamma = -\lim_{A_i \to 0} \sum_{i=1}^{3} \frac{1}{A_i} \frac{\Delta p_i}{\Delta t} = -\lim_{A_i \to 0} \sum_{i=1}^{3} \frac{c}{A_i} \frac{\Delta N_R(x_0)}{\Delta x_0} \cdot \frac{h}{R} = -c \cdot h \lim_{R \to 0} \frac{\Delta\left(\frac{N_R(x_0)}{V_R}\right)}{\Delta x_0}. \text{[22]} \qquad (33)$$

The number of transactions follows by the laws of quantum electrodynamics a Poissonian stochastic process. With a constant average transaction rate, $\varrho_\gamma$[23], the process, is Lorentz-invariant [27]. Hence for the average transaction density $\overline{\lambda}(x_0) = \lim_{R \to 0} \frac{N_R(x_0)}{V_R}$ there holds $\frac{\Delta \overline{\lambda}(x_0)}{\Delta x_0} = \frac{\overline{\lambda}(x_0 + \Delta x_0) - \overline{\lambda}(x_0)}{\Delta x_0} = \varrho_\gamma$ and equation (33) turns into

$$\overline{P}_\gamma = -c \cdot h \cdot \varrho_\gamma. \qquad (34)$$

By defining

$$\Lambda = -\frac{4\pi G}{c^4} \overline{P}_\gamma = \frac{4\pi G h}{c^3} \varrho_\gamma = 4\pi^2 l_P^2 \varrho_\gamma, \qquad (35)$$

---

[20] We assume $E$ to be homogeneously distributed around the origin.
[21] We take the smallest wavelength, compatible with the time-energy inequality.
[22] We set a minus-sign, since the pressure is repulsive. This corresponds to the requirement that $\Lambda$, if traditionally considered as vacuum energy, leads to negative pressure [26].
[23] The transaction rate is the number of transactions per four-volume.

we get from (32) with $\delta_{00}$ denoting the zero-component of the Minkowski metric

$$R_{00} + \Lambda\delta_{00} = \frac{4\pi G}{c^4}T_{00}. \tag{36}$$

If the test systems do not start at rest, $\dot{R}(0) \neq 0$, we have to transform into a local rest frame, such that there is a pressure component originating from the flow of matter, which goes into the equation via the trace of the energy-momentum tensor like in (33). In the local Minkowski metric we have

$$T = tr(T_{\mu\nu}) = T_{00} - \sum_{i=1}^{3} T_{ii}. \tag{37}$$

Adding this component on the right hand side of (36) leaves us finally with

$$R_{00} + \Lambda\delta_{00} = \frac{8\pi G}{c^4}\left(T_{00} - \frac{1}{2}T\delta_{00}\right). \tag{38}$$

Under the assumption of known transformation rules the full Einstein equations

$$R_{\mu\nu} + \Lambda g_{\mu\nu} = \frac{8\pi G}{c^4}\left(T_{\mu\nu} - \frac{1}{2}Tg_{\mu\nu}\right), \tag{39}$$

are equivalent to the fact that equation (38) holds in every local inertial coordinate system around every point in spacetime [25]. Note that the path from equation (29) to (32) and then via (36) to (38) is a path of successive generalization. So, to get back from (38) to Newtonian gravity, we would have to work with masses at low speed, $T \ll T_{00}$, omit $\Lambda$ and then choose weak gravitational fields[24], leading approximately to Euclidean space and equation (29).

We have shown that the entropy increase by a transaction leads to the definition of a natural thermal clock, which measures the rhythm of becoming. Equation (29) synchronizes it with a light-clock, defined by the transferred photon, and a generalized form of (29) turns out to be Einstein's equations. As shown in (33-35) the three-momentum of the exchanged photon leads to the definition of a term $\Lambda$ (35), which represents a cosmological constant with different physical origin than the vacuum energy. The synchronization of thermal clocks from the perspective of different observers is a powerful tool and leads to deviations to Newtonian gravity in the rotation of spiral galaxies, as we show next.

### 4. Corrections to Newtonian Gravity

The observation that the rotation velocity of the outer rims in spiral galaxies does not match the predictions of Newtonian gravity, given the observed mass of these large objects, has lead to the hypothesis of the existence of dark matter. Various proposals as to its true nature have been made, but so far dark matter has remained elusive. Another approach to explain the data has been to make modifications to Newtonian gravity, prominent among which is MOND-theory[25] [18,19]. We will show that the acceleration and velocity found in MOND can be deduced from gauging thermal clocks of different observers, right in the spirit of equation (27). The detailed results can be found in [5] and we refer to this publication, if we do not give all the details here. The starting point is again equation (27)

$$\frac{4}{h}k_B T_{g_R}\frac{c}{g_R} = \frac{1}{\pi^2}. \tag{40}$$

---

[24] In this step we make use of the fact in footnote 14.

[25] Modified Newtonian Dynamics.

After explicitly plugging in formula (28) for the Davies-Unruh temperature $T_{g_R}$ we get the equivalent form, already found in equation (16)

$$T_{g_R} \frac{k_B A_R}{4 l_P^2} = \frac{1}{2} M c^2 = \frac{1}{2} E, \text{[26]} \tag{41}$$

where, as before in equation (14), $A_R = 4\pi R^2$. Equations (40) and (41) will again prove key, this time in a specific spacetime, namely the Schwarzschild-de Sitter solution to Einstein's equations.

*4.1 Schwarzschild-de Sitter spacetime*

To represent space-time with a cosmological constant and in the presence of some mass $M$, with corresponding Schwarzschild radius $R_S = \frac{2GM}{c^2}$, we can use the Schwarzschild-de Sitter solution to Einstein's equations. Its line-element has the form

$$ds^2 = f(r) c^2 dt^2 - \frac{1}{f(r)} dr^2 - r^2 d\Omega^2,$$
$$f(r) = \left(1 - \frac{r^2}{R_0^2} - \frac{R_S}{r}\right), \tag{42}$$

where $R_0 = \sqrt{\frac{3}{\Lambda}}$. Note that the metric (42) is static, although it describes the perspective of a non-inertial observer. In the framework of quantum-events it follows that there is no such thing as a pure vacuum spacetime, since a metric structure and a cosmological term only arise by transactions and the corresponding radiation, which involves and actualizes matter-fields. From a mathematical perspective though, it is of course possible to consider metric (42) as a combination of a vacuum solution to Einstein's equations with metric factor

$$\tilde{f}(r) = \left(1 - \frac{r^2}{R_0^2}\right), \tag{43}$$

called de Sitter space, adjusted by the potential of a gravitating mass $M$, $2\Phi(r) = \frac{2MG}{c^2 r}$. $R_0$ is the Hubble horizon in de Sitter space. Note that the addition of a mass $M$ reduces the horizon $R_0$, $\tilde{f}(R_0) = 0$, to some smaller radius $\hat{R}_0 < R_0$, $f(\hat{R}_0) = 0$. This fact will play a role later on.

In de Sitter space there are the following equalities for the acceleration of the horizon $a_\infty$ [28]

$$a_\infty = c H_0 = \frac{c^2}{R_0} = c^2 \sqrt{\frac{\Lambda}{3}}, \tag{44}$$

where $H_0$ is the Hubble constant. We define $a_0 = \frac{a_\infty}{2}$. For a (hypothetical) acceleration $a(r)$ at radius $r < R_0$ there holds [5]

$$a(r) = \frac{r}{R_0} a_0. \tag{45}$$

Given the fact that in the transactional ontology a metric emerges from interacting matter, a pure vacuum spacetime and therefore an acceleration of empty space, $a(r)$, are meaningless. The factor $\frac{r}{R_0}$ could, however, more meaningfully be attributed to the entropy expression $S(r) = \frac{k_B A_r}{4 l_P^2}$ in (41) in order to define the de Sitter entropy $S_{dS}(r)$

---

[26] Note that we have multiplied both sides with a factor $\frac{1}{2}$ for convention's sake.

$$S_{dS}(r) = \frac{r}{R_0} S(r). \tag{46}$$

For different reasons[27], the same definition arises in [29], where the horizon-entropy

$$S(R_0) = \frac{k_B A_{R_0}}{4 l_P^2} \tag{47}$$

is supposed to be equally distributed over the bulk of de Sitter space and where at $t = 0$, say, a ball around the origin with radius $r$ and volume $V(r)$ contains de Sitter entropy

$$S_{dS}(r) = \frac{V(r)}{V_0}. \tag{48}$$

$V_0$ is chosen[28] in such a way that there results again equation (46)

$$S_{dS}(r) = \frac{r}{R_0} S(r). \tag{49}$$

*4.2 Effective entropy in Schwarzschild- de Sitter spacetime*

Since, as mentioned above, the mass-induced potential $2\Phi(r)$ in (42) reduces the horizon, $\hat{R}_0 < R_0$, it also reduces the corresponding entropy (47). In this paragraph we follow the exposition in [29]. For $|\Phi(r)| \ll 1$ the horizon changes through addition of mass by a negative amount of

$$\Delta R_0 = \Phi(R_0) \cdot R_0. \tag{50}$$

Hence the horizon-entropy changes under $R_0 \to R_0 + \Delta R_0$ to first order by

$$\Delta S_{dS}(R_0) = \Delta R_0 \frac{d}{dR_0} \left( \frac{k_B A_{R_0}}{4 l_P^2} \right) = -\frac{2\pi c k_B M R_0}{\hbar}. \tag{51}$$

In order to calculate first order entropy change in regions far away from the horizon, $\frac{r}{R_0} \ll 1$, we take in the presence of a mass $M$ the derivative of $S(r)$ with respect to the geodesic distance $ds = (1 + \Phi(r))^{-1} dr$, whereas in vacuum this distance is simply $ds = dr$. The first order difference of entropy in the two situations becomes

$$\frac{d}{ds} \left( \frac{k_B A_r}{4 l_P^2} \right) \bigg|_{M=0}^{M \neq 0} = \Phi(r) \frac{d}{dr} \left( \frac{k_B A_r}{4 l_P^2} \right) = -\frac{2\pi c k_B M}{\hbar}. \tag{52}$$

The right hand side of (52) is the amount of entropy, which a mass $M$ takes away from the entropy of a spherical region of radius $r$. In view of (51) we define the infinitesimal change $\frac{\Delta S_{dS}(r)}{dr}$ to be differentiation with respect to coordinate $r$ only, to get after integration

$$\Delta S_{dS}(r) = -\frac{2\pi c k_B M}{\hbar} r. \tag{53}$$

---

[27] The arguments are of string-theoretic type.

[28] $V_0 = \frac{4 G \hbar R_0}{3 c^2}$.

Multiplying equation (53) both in the nominator and denominator by $R_0$ and remembering (51) leads to the form

$$\Delta S_{dS}(r) = \frac{r}{R_0} \Delta S_{dS}(R_0). \tag{54}$$

We now define the effective de Sitter entropy after addition of a mass $M$, $\bar{S}_{dS}(r)$, by

$$\bar{S}_{dS}(r) = S_{dS}(r) + \Delta S_{dS}(r) = \frac{k_B r}{R_0}\left(\frac{A_r}{4l_P^2} - \frac{2\pi c M R_0}{\hbar}\right). \tag{55}$$

We are now ready to tackle the main goal of this section, namely to calculate the effective acceleration $\bar{a}(r)$ in Schwarzschild-de Sitter spacetime.

*4.3 Effective acceleration: the main formula*

The effective acceleration $\bar{a}(r)$ in Schwarzschild-de Sitter spacetime describes the point of view of a non-inertial observer for whom the flow of matter is subject to a combined vacuum-gravity acceleration. At the same time we can also take the viewpoint of an observer co-moving with the vacuum expansion and only experiencing gravitative pull $g(r)$. Both accounts have to be consistent. In the presence of matter, equation (45) makes sense again for $\bar{a}(r)$ and should approximately hold in regions, where gravitational acceleration is sufficiently small

$$\bar{a}(r) \sim \frac{r}{R_0} a_0. \tag{56}$$

We can say much more about $\bar{a}(r)$ by noticing that for both observers there hold equations (40) and (41) or their analogue, since they both locally define thermal clocks, which they can gauge by a light clock (27). By the local invariance of the speed of light we need for consistency reason in the range $r > r_0$, $r_0$ to be determined later, [6],

$$T_{\bar{a}(r)} \bar{S}_{dS}(r) = T_{g(r)} S(r). \tag{57}$$

Equation (57) is the key relation and by (55) it turns into

$$T_{\bar{a}(r)}\big(S_{dS}(r) + \Delta S_{dS}(r)\big) = T_{g(r)} S(r). \tag{58}$$

By applying (49), (53) and (56) we get

$$\frac{\bar{a}(r)^2}{a_0} S(r) - \bar{a}(r)|\Delta S_{dS}(r)| - g(r) S(r) = 0. \tag{59}$$

For fixed $r \geq r_0$ (59) is quadratic in $\bar{a}(r)$ with one positive solution

$$\bar{a}(r) = a_0 \frac{\left(\left(\frac{|\Delta S_{dS}(r)|}{S(r)}\right) + \sqrt{\left(\frac{|\Delta S_{dS}(r)|}{S(r)}\right)^2 + \frac{4g(r)}{a_0}}\right)}{2}. \tag{60}$$

The Newtonian potential can be written $|\Phi(r)| = \frac{1}{2}\frac{|\Delta S_{dS}(r)|}{S(r)}$ and hence (60) turns into

$$\overline{a}(r) = a_0 \frac{\left(2|\Phi(r)| + \sqrt{4|\Phi(r)|^2 + \frac{4|\Phi(r)|^2 c^4}{MGa_0}}\right)}{2}. \tag{61}$$

After factoring out we finally get the main formula

$$\overline{a}(r) = a_0|\Phi(r)|\left(1 + \sqrt{1 + \frac{c^4}{MGa_0}}\right). \tag{62}$$

*4.4 Discussion of the main formula*

We already mentioned that equation (54) is valid for radii, $r \geq r_0$, bigger than some limit radius $r_0$. This limit radius follows immediately from equation (55), because for (57) to make sense there must hold

$$\overline{S}_{dS}(r) > 0. \tag{63}$$

This means by (55)

$$\frac{A_r}{4l_P^2} > \frac{2\pi c M R_0}{\hbar}, \tag{64}$$

and hence

$$r > r_0 = \sqrt{\frac{MG}{a_0}}. \tag{65}$$

Below this limit $\overline{S}_{dS}(r)$ is exhausted, gravity dominates and there holds the Newtonian regime. We still need to check consistency at $r = r_0$. To do this we make use of the fact that by (55) we have $S(r_0) = \Delta S_{dS}(R_0)$ and hence $|\Phi(r_0)| = \frac{r_0}{2R_0}$. Since $\frac{r_0}{R_0} a_0 \ll 1$[29], equation (62) turns by (44) into

$$\overline{a}(r_0) \approx \frac{r_0}{2R_0} a_0 \sqrt{\frac{c^4}{MGa_0}} = \sqrt{\frac{c^4}{4R_0^2}} = a_0. \tag{66}$$

On the other hand we get by a direct calculation

$$g(r_0) = \frac{c^2|\Phi(r_0)|}{r_0} = \frac{c^2}{2R_0} = a_0. \tag{67}$$

Equations (66) and (67) guarantee consistency of the regimes at $r = r_0$.

At the same time, since $|\Phi(r)|a_0 \ll 1$ for $r > r_0$, we can simplify equation (62) further to get

$$\overline{a}(r) \approx |\Phi(r)|a_0 \sqrt{\frac{c^4}{MGa_0}} = \frac{\sqrt{MGa_0}}{r}. \tag{68}$$

---

[29] $a_0 \sim 10^{-10} \frac{m}{s^2}$.

The right hand side of (68) is exactly the expression suggested in the original MOND-theory at accelerations above $\sim a_0$ [15]. Via the link $\bar{a}(r) = \frac{v^2(r)}{r}$ we get

$$v^2 = \sqrt{MGa_0}, \qquad (69)$$

which fits observed velocity-data in spiral galaxies.

## 5. Conclusion

In our work gravity reveals itself to be the result of entropic effects in connection with electromagnetically induced transactions. Gravitation is therefore a consequence of coming-into-being (at the empirical level) *per se*, and hence universal. The photons involved in transactions serve as naturally inbuilt clocks, which measure the rhythm of becoming, and Einstein's equations turn out to result from synchronizing these clocks, as they appear from specific perspectives, by the invariant speed of light. Effects, attributed to "dark matter", are also shown to stem from this synchronizing process, while "dark energy" expansion is a consequence of the exchange of photon-momentum in transactions. Since electromagnetism is the root of these transactions, it is clear that gravity can at most spread at the speed of light and that the "charge" of gravity is inertial mass. Hence there holds the equivalence of inertial and heavy mass. Our approach, together with a result in [24], also shows that the length of a timelike curve is a good representation for the time which elapses by succeeding transactions. Therefore relativistic proper-time indeed measures any experienced time-flow, biological, mechanical or other, since it measures the rhythm of coming-into-being of every individual physical system (more precisely, the sets of actualized events associated with such systems) in the universe. Our approach lends itself to the relational view regarding the nature of spacetime [17] and the derivation of gravity also indicates that, while being a consequence of quantum mechanics, it is itself not described by some quantum field and its interactions.[30] While of course, elementary particles do have mass, the mathematics in 2.4. gives hints that at very small distances, e.g. below the Compton radius in Bekenstein's approach, there are no transactions anymore, which would lead to no manifestation of gravity in our approach. In this context it should also be noted that due to conservation laws, transactions can only occur between bound states subject to internal energy levels, which do not attend elementary particles [1]. We state this fact as an observation, which has otherwise no significance to the results in this paper.

We don't know, at present, whether transactions describe correctly what nature actually does, and there are consequently different approaches to explain gravity. The beauty of our approach lies in the fact that gravity, including dark energy and dark matter, can very elegantly be derived as a consequence of the existing formalism of quantum physics together with principles of thermodynamics. All that is needed is a different perspective as to what quantum physics tells us that nature really does, and a number of open questions surrounding gravity naturally find an answer.

References.

---

[30] This does not imply that spacetime is not discrete, which it probably is, as discussed in section 1.